\documentclass[lettersize,journal]{IEEEtran}

\usepackage{amsmath,amsfonts,amssymb}
\usepackage{bm}
\usepackage{textcomp}
\usepackage{stfloats}
\usepackage{url}
\usepackage{verbatim}
\usepackage{graphicx}
\usepackage[shortlabels]{enumitem}
\usepackage{needspace}
\usepackage{balance}
\usepackage{comment}
\usepackage[table]{xcolor}
\usepackage{algorithm}
\usepackage{algpseudocode}
\usepackage{wasysym}
\usepackage{multirow}
\usepackage[numbers,sort&compress]{natbib}

\usepackage{blindtext}
\usepackage{wrapfig}
\usepackage{booktabs}
\usepackage{subfigure}
\usepackage{tcolorbox}
\tcbuselibrary{breakable}
\usepackage{soul}
\usepackage{colortbl}
\usepackage{listings}
\usepackage{array}
\usepackage{adjustbox}
\usepackage{seqsplit}
\usepackage{xspace}
\usepackage{hyperref}
\usepackage[hyphenbreaks]{breakurl}

\definecolor{lightpink}{rgb}{1.0, 0.87, 0.87}
\definecolor{lightpurple}{rgb}{0.94, 0.85, 0.94}
\definecolor{lightgreen}{rgb}{0.88, 1.0, 0.88}
\definecolor{lightyellow}{rgb}{1.0, 1.0, 0.88}
\definecolor{lightblue}{rgb}{0.88, 0.94, 1.0}
\definecolor{lightgrey}{rgb}{0.8, 0.8, 0.8}
\definecolor{codegreen}{rgb}{0,0.6,0}
\definecolor{codegray}{rgb}{0.5,0.5,0.5}
\definecolor{codepurple}{rgb}{0.58,0,0.82}
\definecolor{backcolour}{rgb}{0.95,0.95,0.92}

\lstdefinestyle{mystyle}{
    backgroundcolor=\color{backcolour},
    commentstyle=\color{codegreen},
    keywordstyle=\color{magenta},
    numberstyle=\tiny\color{codegray},
    stringstyle=\color{codepurple},
    basicstyle=\footnotesize\ttfamily,
    breakatwhitespace=false,
    breaklines=true,
    captionpos=b,
    keepspaces=true,
    numbers=left,
    numbersep=5pt,
    showspaces=false,
    showstringspaces=false,
    showtabs=false,
    tabsize=3,
    frame=tb,
    aboveskip=2mm,
    belowskip=2mm
}
\newcommand{\SymLift}{\textsc{SymLift}\xspace}

\begin{document}

\title{\SymLift: Neural Semantic Lifting for Compile-Free Vulnerability Detection in Context-Incomplete Code Fragments}

\author{Xiarun Chen, Jiaxing Xu, Bozhi Wu, Chenhui Cui, Chunrong Fang~\IEEEmembership{Member,~IEEE}, Weiping Wen, Tao Li, Rubing Huang,~\IEEEmembership{Senior Member,~IEEE}, Tsong Yueh Chen,~\IEEEmembership{IEEE Fellow,~IEEE}

\thanks{Xiarun Chen, Jiaxing Xu, and Weiping Wen are with Peking University, Beijing 100191, China (e-mail: jxxu25@stu.pku.edu.cn; xiar\_c@pku.edu.cn; weipingwen@pku.edu.cn).}
\thanks{Bozhi Wu is with Nanyang Technological University, Singapore 639798, Singapore (e-mail: bozhi001@e.ntu.edu.sg).}
\thanks{Chenhui Cui is with the University of Alberta, Edmonton T6G 2R3, Canada (e-mail: chenhui.cui@ualberta.ca).}
\thanks{Chunrong Fang is with Nanjing University, Nanjing 210023, China (e-mail: fangchunrong@nju.edu.cn).}
\thanks{Tao Li is with Beihang University, Beijing 100191, China (e-mail: abelli@buaa.edu.cn).}
\thanks{Rubing Huang is with Macau University of Science and Technology, Macao SAR 999078, China (e-mail: rbhuang@must.edu.mo).}
\thanks{Tsong Yueh Chen is with Swinburne University of Technology, Melbourne 3122, Australia (e-mail: tychen@swin.edu.au).}
\IEEEcompsocitemizethanks{\IEEEcompsocthanksitem Xiarun Chen and Jiaxing Xu are co-first authors and contributed equally to this work.}
\IEEEcompsocitemizethanks{\IEEEcompsocthanksitem Bozhi Wu and Tao Li are corresponding authors.}
}

\maketitle

\begin{abstract}
Vulnerability detection is increasingly performed on context-incomplete code fragments, such as code generated by large language models (LLMs) and individual functions extracted from larger repositories.
Missing project context can prevent build-dependent analyzers from constructing the control-flow or data-flow relations required for vulnerability detection.
Learning-based detectors and LLMs used directly as vulnerability detectors can analyze such fragments, but their predictions alone do not establish an auditable derivation from explicit program relations.
To enable auditable vulnerability detection without requiring the input to compile, we present \SymLift, a framework centered on \emph{Vulnerability-Oriented Neural Semantic Lifting}.
\SymLift uses a neural lifter to reconstruct vulnerability-relevant control-flow, data-flow, memory, and resource relations in a constrained LLVM-like intermediate representation (IR), without reconstructing the full project context or predicting vulnerability labels directly.
A deterministic parser validates the IR and maps its supported operations and relations into normalized evidence facts.
A Datalog engine applies explicit vulnerability rules to derive findings with supporting facts and derivation traces.
\SymLift keeps vulnerability reasoning deterministic for fixed facts and rules.
We conduct experiments on FormAI, Juliet, and MegaVul.
When complete Juliet programs are reduced to function-level fragments, the Clang compilation success rate drops from 100\% to 12.00\%, whereas \SymLift's analysis completion rate decreases by only 1.17 percentage points.
After task-specific fine-tuning, the neural lifter achieves a control-flow intersection over union (IoU) of 96.85\% on FormAI.
The task-fine-tuned \SymLift-Qwen-7B configuration achieves a Target-F1 of 94.41\% on FormAI and, across six vulnerability families on MegaVul, 
a micro-averaged precision of 98.24\% and a micro-averaged Target-F1 of 91.66\%.
Overall, vulnerability-oriented neural semantic lifting supports accurate and auditable detection in the evaluated context-incomplete settings, even when the reconstructed program relations are partial.
\end{abstract}

\begin{IEEEkeywords}
Vulnerability detection, LLM, static analysis, symbolic reasoning
\end{IEEEkeywords}

\section{Introduction}
\label{SEC:Introduction}

\IEEEPARstart{V}{ULNERABILITY} analysis increasingly encounters code fragments that must be assessed without their project context.
For example, functions generated by large language models (LLMs) may contain security vulnerabilities and lack necessary defensive checks~\cite{pearce2025asleep,majdinasab2024assessing,fu2023security,tihanyi2025secure}.
Recent studies report security weaknesses in C/C++ generation~\cite{shahid2026llm} and code from developer--LLM interactions~\cite{belozerov2026secure}.
When these outputs are analyzed in isolation, context required for vulnerability analysis may be unavailable~\cite{zhao2025towards}.
Similarly, individual functions extracted from larger repositories~\cite{fan2020ac,ni2024megavul} may be analyzed without their original project context.
These fragments can be context-incomplete even if they are syntactically valid and can compile independently~\cite{vero2025baxbench,peng2025cweval}.
We refer to code fragments that lack some context required for conventional program analysis as \emph{context-incomplete code fragments}.

Traditional static analyzers derive vulnerability-relevant evidence from typed abstract syntax trees (ASTs), control-flow graphs (CFGs), and definition-use relations~\cite{codeql,clang-static-analyzer,cppcheck}.
When context is missing, compilers may fail to construct these representations, while tolerant parsers may recover only partial syntactic structure~\cite{Semgrep2021}.
However, recovering an AST alone does not ensure that relations required by vulnerability rules can be recovered.
We refer to the mismatch between available source code and the recovery of program relations needed for downstream vulnerability detection as the \emph{compilation gap}.
Under missing context, this gap can manifest as failed relation construction, incomplete recovery, or structural divergence from representations obtained with more complete context.

Learning-based vulnerability detectors can reduce compilation dependence by encoding source-code tokens, program slices, or learned graph representations~\cite{li2018vuldeepecker,li2021sysevr,feng2020codebert,guo2020graphcodebert,wang2021codet5,guo2022unixcoder}.
However, their probabilistic predictions are difficult to trace to explicit program evidence~\cite{chakraborty2024revisiting,ding2024vulnerability,li2026revisiting,safdar2026data}, and their detection performance can degrade under realistic data distributions~\cite{li2026out}.
Moreover, LLMs have been used directly as vulnerability detectors~\cite{zhang2024prompt,du2026vul, peng2026three, zhu2026vulinstruct}.
Their predictions and explanations alone do not establish an auditable derivation from explicit program relations.
IRIS~\cite{li2024iris} uses LLMs to infer taint specifications for project-level CodeQL reasoning.
MoCQ~\cite{li2025automated} uses LLMs to generate and iteratively refine vulnerability queries executed over Joern and CodeQL representations.
Their downstream reasoning still relies on analysis frontends to construct the required program representations.
These limitations highlight the need to reconstruct vulnerability-relevant program representations from context-incomplete code fragments and support deterministic, auditable vulnerability reasoning.

To enable auditable vulnerability detection on context-incomplete code fragments, we propose \SymLift, a compile-free neuro-symbolic framework.
Here, \emph{compile-free} means that, at analysis time, \SymLift requires neither a complete build environment nor successful compilation of the input code.
The core of \SymLift is \emph{Vulnerability-Oriented Neural Semantic Lifting}, which uses a fine-tuned LLM to reconstruct vulnerability-relevant program relations in a constrained LLVM-like intermediate representation (IR).
The neural lifter neither reconstructs the full project context nor directly predicts vulnerability labels.

\SymLift uses LLVM-like IR as the interface in a three-stage pipeline: (1) the neural lifter generates IR (Stage I); (2) a deterministic parser validates it and maps supported operations and relations into evidence facts (Stage II); and (3) \SymLift evaluates explicit vulnerability rules using Souffl\'e Datalog~\cite{jordan2016souffle} (Stage III).
Each reported vulnerability is linked to supporting facts, a matched rule, and a derivation trace.

\SymLift's core semantic-lifting objective is supported by three design choices: (1) The neural lifter focuses on vulnerability-relevant relations.
(2) LLVM-like IR provides a constrained semantic interface.
(3) The deterministic parser and explicit rules separate neural semantic reconstruction from vulnerability judgment.
The IR layer does not eliminate uncertainty in neural lifting; it makes the reconstructed relations inspectable and supports deterministic reasoning for fixed facts and rules.

We implement and evaluate \SymLift on FormAI, Juliet, and MegaVul~\cite{formai,juliet,ni2024megavul}.
On Juliet, reducing programs to function-level fragments lowers compilation success from 100\% to 12.00\%, while \SymLift's analysis completion drops by only 1.17 percentage points.
On FormAI, \SymLift-Qwen-7B achieves 87.83\% overall relation-level intersection over union (IoU), 96.85\% control-flow IoU, and 94.41\% Target-F1, exceeding the strongest direct-LLM baseline, Gemini-3.5-Flash, by 24.08 percentage points.
On MegaVul, it achieves 98.24\% precision and 91.66\% micro-averaged Target-F1.

In summary, this paper makes the following contributions:
\begin{enumerate}[leftmargin=2em, itemsep=0pt, parsep=0pt, topsep=0.5pt, partopsep=0pt]
\item We characterize the compilation gap for context-incomplete code fragments and show that syntactic recoverability does not guarantee the availability of vulnerability-relevant control-flow and data-flow relations.

\item We develop \emph{Vulnerability-Oriented Neural Semantic Lifting}, which uses two-stage fine-tuning to reconstruct vulnerability-relevant program relations in constrained LLVM-like IR.

\item We design a deterministic IR-to-fact parser and an explicit vulnerability-rule system that separate evidence extraction from vulnerability judgment and produce reproducible detection results with supporting facts and derivation traces for fixed input facts and rules.

\item We implement \SymLift and evaluate representation availability, semantic fidelity, and vulnerability-detection effectiveness on AI-generated code, controlled context removal, and real-world
CVE-related functions.\footnote{The implementation and experimental data are publicly available at
\url{https://figshare.com/s/cda7af5b537730fd7ab3}.}

\end{enumerate}

The remainder of this paper is organized as follows.
Section~\ref{SEC:Background} introduces the technical background.
Section~\ref{SEC:Motivation} illustrates the challenges of detecting vulnerabilities in context-incomplete code fragments.
Section~\ref{SEC:Approach} presents the design of \SymLift.
Section~\ref{SEC:EmpiricalStudy} reports the empirical evaluation.
Section~\ref{SEC:Discussion} discusses practical implications and limitations.
Sections~\ref{SEC:Related} and~\ref{SEC:ThreatsToValidity} review related work and threats to validity, respectively.
Finally, Section~\ref{SEC:Conclusion} concludes the paper.

\section{Background}
\label{SEC:Background}

This section introduces the technical foundations underlying \SymLift.

\subsection{Context-Incomplete Code and Program Representations}
\label{SEC:BackgroundContext}

A \emph{context-incomplete} fragment lacks some surrounding information needed for semantic analysis, as in LLM-generated code~\cite{formai,vangala2025ai} and function-level datasets~\cite{fan2020ac,ni2024megavul}.
Syntactic validity does not ensure compilation, and compilable fragments may still lack external implementations; \emph{independently non-compilable} means the supplied fragment cannot compile outside its original project.
CodeQL~\cite{codeql}, Clang Static Analyzer~\cite{clang-static-analyzer}, and Cppcheck~\cite{cppcheck} use structured representations~\cite{lipp2022empirical}; successful parsing alone does not ensure the required relations.

\subsection{LLVM Intermediate Representation}
\label{SEC:BackgroundIR}

LLVM IR is typed and instruction-level~\cite{LLVM10.1145}, with one definition per virtual register in static single assignment (SSA) form~\cite{llvm-langref}.
Branches express control flow; \texttt{load}/\texttt{store} access memory, whereas \texttt{getelementptr} computes addresses without accessing memory.
Previous work uses LLVM IR for Datalog-based vulnerability analysis~\cite{schilling2022vandalir}.
Our \emph{LLVM-like IR} follows LLVM conventions within a restricted, parser-supported subset, but generated outputs may be malformed.
\texttt{llvm-as} validates reference IR and, with the parser, sampled IR during structural training; inference extracts facts without requiring assembly acceptance.
Analysis completion does not establish semantic equivalence.

\subsection{Fact-Based Vulnerability Reasoning with Datalog}
\label{SEC:BackgroundDatalog}

Datalog supports static analyses~\cite{jordan2016souffle} through input facts and rules defining derived relations.
A rule may combine operational evidence with the absence of derivable safety conditions; fixed facts and rules yield deterministic results that can be traced to supporting evidence.
Souffl\'e executes these analyses~\cite{jordan2016souffle}, but can reason only over available input relations.

\section{Motivating Example}
\label{SEC:Motivation}

As shown in Listing~\ref{lst:motivating-example}, the fragment contains a potentially unsafe array access, but the surrounding declarations and type definitions needed for conventional analysis are unavailable.

\begin{lstlisting}[
    language=C,
    caption={A context-incomplete function with a potential negative-index access.},
    label={lst:motivating-example},
    basicstyle=\scriptsize\ttfamily,
    frame=single,
    breaklines=true,
    numbers=left,
    numberstyle=\tiny,
    stepnumber=1,
    numbersep=5pt,
    showstringspaces=false,
    tabsize=2
]
int read_entry(Context *ctx, int idx) {
    int count = get_entry_count(ctx);
    if (idx < count)
        return ctx->entries[idx];
    return -1;
}
\end{lstlisting}

To isolate the index-bounds issue, we assume that \texttt{ctx->entries} denotes the base of a valid array, and \texttt{get\_entry\_count(ctx)} returns its number of accessible elements.
These assumptions are used only to explain the example; they are not analyst-supplied inputs to \SymLift.
No caller-side non-negativity condition is assumed.

Under these assumptions, \texttt{idx < count} establishes an upper bound but not the lower bound \texttt{idx >= 0}.
A negative value may therefore reach \texttt{ctx->entries[idx]}, causing a potential access before the first array element.
The example therefore illustrates the locally visible missing-lower-bound condition without establishing its reachability across the complete program's calling contexts.

As written, the fragment cannot be compiled independently because the definition of \texttt{Context} and the declaration of \texttt{get\_entry\_count} are unavailable.
A tolerant parser may recover syntax without establishing the required typed field access, definition-use relations, or control dependence.
Build-dependent analyzers may therefore fail to construct, or only partially construct, the program representations required by their vulnerability rules~\cite{codeql,clang-static-analyzer}.

\SymLift bridges this gap without asking the neural lifter to predict a vulnerability label directly.
For this example, the generated LLVM-like IR represents \texttt{idx < count} with \texttt{icmp slt} and \texttt{br}, and the array access with \texttt{getelementptr} and \texttt{load}, using the same SSA value \texttt{idx}.
The deterministic parser maps these instructions to Base Facts and derives the corresponding \texttt{DefUse}, \texttt{ControlDependent}, and \texttt{ArrayAccess} relations.
Given the recovered indexed-access and upper-bound evidence, the decision rule emits \texttt{missing\_bounds\_check} if \texttt{CompleteBoundsGuard} cannot be derived.
The report is therefore traceable to the reconstructed instructions, extracted Base Facts, derived relations, and matched decision rule.

The next section presents the complete pipeline of our approach.

\section{Approach}
\label{SEC:Approach}

This section describes the design of \SymLift.

\subsection{Overview}
\label{SEC:ApproachOverview}

\begin{figure*}[t]
    \centering
    \includegraphics[width=0.95\textwidth]{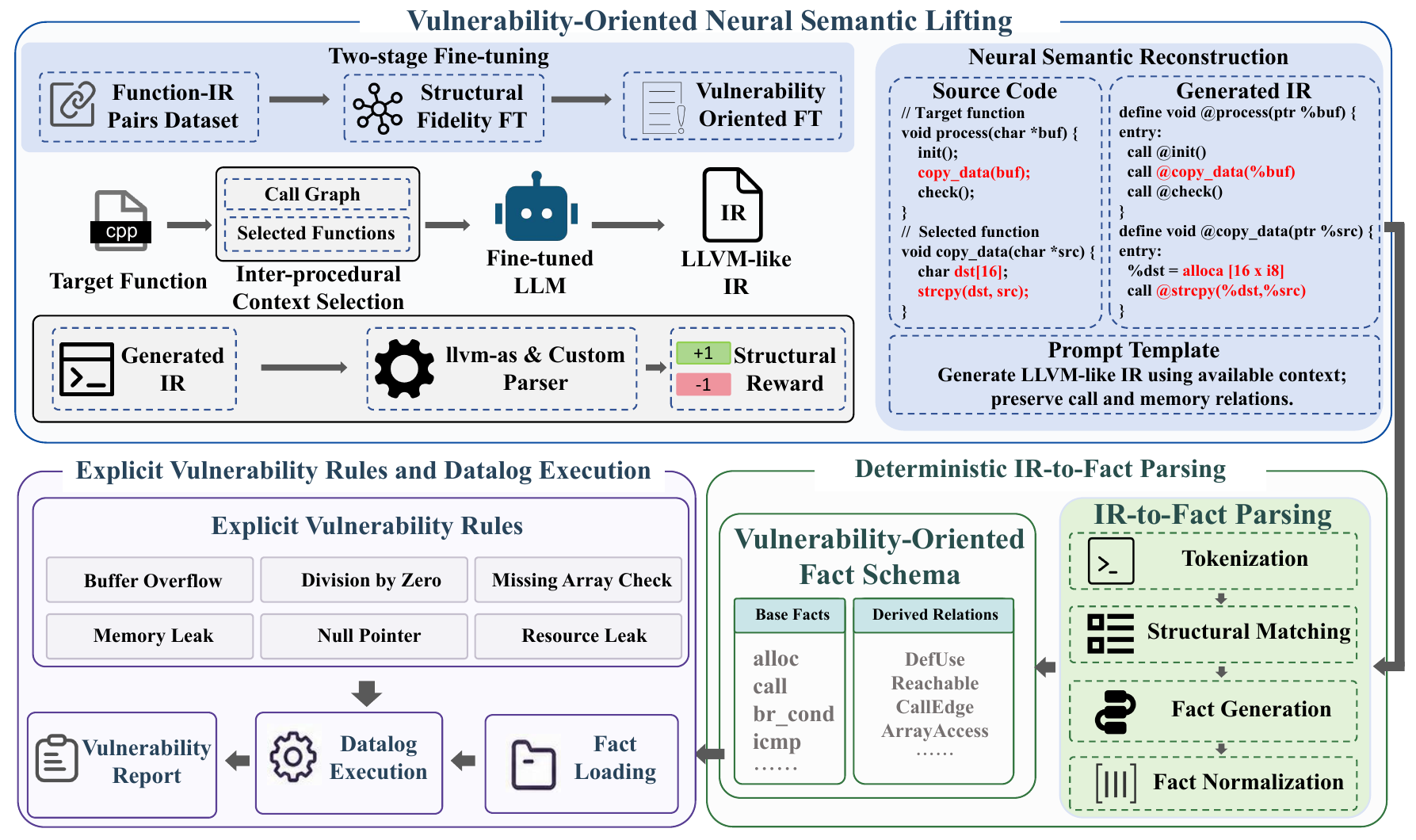}
    \caption{Overview of the \SymLift workflow.}
    \label{fig:architecture}
\end{figure*}

Figure~\ref{fig:architecture} illustrates the three-stage pipeline.
Stage~I lifts context-incomplete functions into LLVM-like IR; Stage~II parses IR and constructs normalized evidence facts; Stage~III evaluates explicit vulnerability rules using Souffl\'e Datalog~\cite{jordan2016souffle}.

Direct fact generation can also be schema-constrained; the IR route first reconstructs an instruction-level representation from which the parser extracts Base Facts.
Its typed operations, basic blocks, branch targets, pointer computations, and static single-assignment (SSA) form make value dependencies explicit and permit structural validation before rule evaluation.
Table~\ref{tab:rq3-direct-facts} evaluates this route against direct fact generation.

\subsection{Stage I: Vulnerability-Oriented Neural Semantic Lifting}
\label{SEC:StageI}

Stage~I receives a context-incomplete code fragment and produces LLVM-like IR designed to preserve the program relations represented by the downstream Fact Schema.

\subsubsection{Analysis Granularity and Preprocessing}

The target function is the basic unit of analysis.
With project sources available, Tree-sitter builds direct-call edges only for unique project-local definitions, preferring same-file \texttt{static} definitions.
Bidirectional breadth-first traversal from the target retains functions within $d=3$ call edges, ordered by distance and repository-relative source location, up to 50 including the target.
Indirect, ambiguous, library, or unavailable calls add no traversal edges; their call-site information remains available for IR lifting, with unresolved function-pointer calls marked indirect.
Without project sources, analysis uses only the target; otherwise, selected functions are lifted independently and their facts analyzed jointly, with auxiliary-only findings excluded.
Training and inference share function slicing by source ranges and recursive \texttt{typedef}/\texttt{struct}/\texttt{union}/\texttt{enum} dependency selection, retaining complete types and field order.
Training pairs slices with Clang reference IR stripped of unrelated metadata; inference uses no Clang or reference IR, and unavailable context is omitted rather than fabricated.

\subsubsection{Two-Stage Fine-Tuning}

\begin{figure}[t]
    \centering
    \includegraphics[width=0.5\textwidth]{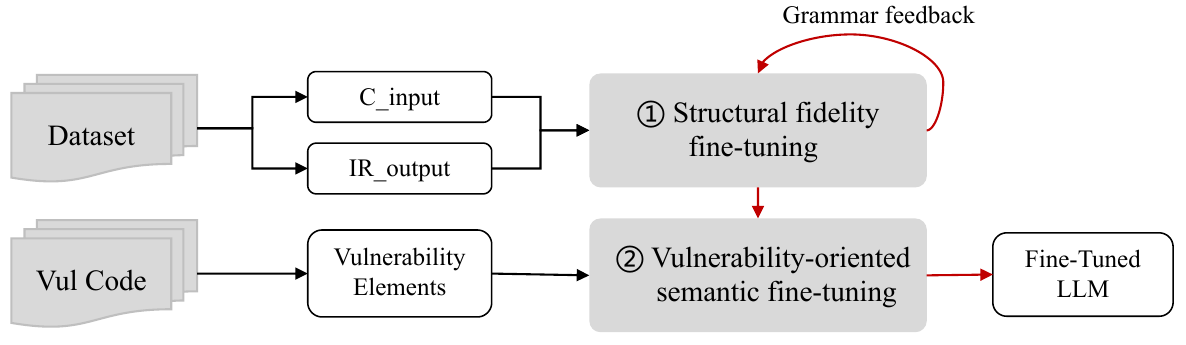}
    \caption{Two-stage fine-tuning for Vulnerability-Oriented Neural Semantic Lifting.}
    \label{fig:two-stage-finetuning}
\end{figure}

As shown in Figure~\ref{fig:two-stage-finetuning}, we employ two consecutive fine-tuning stages for structural conformity and vulnerability-relevant semantic preservation.

\textbf{Structural-fidelity fine-tuning.}
The model is first fine-tuned on paired C functions and Clang-generated IR.
Fine-tuning prompts constrain function definitions and types, entry blocks, instruction ordering, block termination, branch targets, and operand-result consistency.
During structural-fidelity training, token-level cross-entropy against the reference IR is supplemented with structural feedback from four sampled outputs per input, checked by both \texttt{llvm-as} and the custom parser.
Because these validators are non-differentiable, their feedback is incorporated through a policy-gradient objective:
\begin{equation}
\nabla_{\theta}\mathcal{L}_{\mathrm{struct}} = -\mathbb{E}_{Y\sim P_{\theta}} \left[ A(Y) \nabla_{\theta}\log P_{\theta}(Y\mid X) \right],
\label{eq:syntax-loss}
\end{equation}
where $X$ denotes the input function and $Y$ denotes a sampled IR output.
The structural reward is defined as:
\begin{equation}
R_{\mathrm{struct}}(Y)=
\begin{cases}
1, &
\begin{aligned}[t]
&\text{if both \texttt{llvm-as} and the}\\
&\text{custom IR parser accept }Y,
\end{aligned}
\\
-1, & \text{otherwise.}
\end{cases}
\label{eq:parser-reward}
\end{equation}
This reward encourages LLVM assembly validity and compatibility with the downstream parser, rather than establishing semantic equivalence with the source function.
For each candidate, $A(Y)$ is obtained by normalizing and clipping its reward minus the mean reward of the other three outputs to $[-2,2]$.
The final training objective combines IR reconstruction supervision and structural feedback:
\begin{equation}
\mathcal{L}_{\mathrm{total}} = \mathcal{L}_{\mathrm{CE}} + \alpha\mathcal{L}_{\mathrm{struct}},
\label{eq:total-loss}
\end{equation}
where $\alpha=0.1$.
Both terms are optimized jointly; when all four rewards are identical, only the cross-entropy contribution is retained for that input.

\textbf{Vulnerability-oriented semantic fine-tuning.}
The second stage uses the security-oriented FormAI training subset together with matched non-vulnerable functions to preserve target-family relations, including bounds and zero checks, pointer accesses and NULL checks, allocation-release lifecycles, error paths, and memory-sensitive API calls.
Data augmentation applies semantics-preserving variable renaming, reordering of dependence-independent statements, and equivalent source-level rewrites to the training functions.
Each of the 12,980 deduplicated functions receives one augmented variant re-aligned with its original reference IR; rejecting 214 variants whose re-aligned IR fails the custom parser's structural checks leaves 25,746 training records.
The model initialization, adapter configuration, and optimization settings are reported in Section~\ref{SEC:TrainingConfiguration}.

\subsubsection{Relation-Oriented Model Validation}

Project-level validation uses 1000 functions from zlib 1.2.11 and libpng 1.6.37, excluded from both training corpora and built with the structural-fidelity toolchain and flags.
Both projects are fixed before training; 500 functions per project are selected by SHA-256 ranking of seed 42 concatenated with stable identifiers.
Generated and reference IR are normalized into operation types, definition-use edges, branches, and control-flow transitions rather than compared as raw text.
Model selection is based on structural validity, relation-level agreement, vulnerability-relevant relation preservation, and generation latency.
The held-out set is used for model selection, never for gradient updates.

\subsection{Stage II: Fact Schema Design and Deterministic IR-to-Fact Parsing}
\label{SEC:StageII}

Stage~II validates the candidate LLVM-like IR and deterministically converts its supported content into normalized evidence facts, leaving vulnerability judgments to Stage~III.

\subsubsection{Vulnerability-Oriented Fact Schema}

The Vulnerability-Oriented Fact Schema defines a stable interface between LLVM-like IR and explicit vulnerability rules.
It preserves program relations as typed predicates and normalizes equivalent IR relations into the same vocabulary.
New evidence relations and rules can be introduced without retraining the neural lifter when the generated IR already contains the required information.

\textbf{Base Facts} map supported IR instructions to typed predicates; \textbf{Derived Program Relations} are constructed mechanically from these facts without asserting vulnerabilities.
Table~\ref{tab:fact-mapping} summarizes their predicates, semantics, and related rule families.

\begin{table*}[t]
\centering
\caption{Evidence relations represented by the Vulnerability-Oriented Fact Schema.}
\label{tab:fact-mapping}
\footnotesize
\renewcommand{\arraystretch}{1.15}
\setlength{\tabcolsep}{5pt}
\begin{tabular}{@{}>{\raggedright\arraybackslash}p{0.14\textwidth}>{\raggedright\arraybackslash}p{0.28\textwidth}>{\raggedright\arraybackslash}p{0.30\textwidth}>{\raggedright\arraybackslash}p{\dimexpr0.28\textwidth-6\tabcolsep\relax}@{}}
\toprule
\textbf{Evidence Type} & \textbf{Representative Predicates} & \textbf{Captured Semantics} & \textbf{Related Rule Families} \\
\midrule
Control flow & \texttt{icmp}, \texttt{br\_cond}, \texttt{Reachable}, \texttt{ControlDependent} & Comparisons, guarded paths, and basic-block reachability & Bounds, NULL checks, and division-by-zero rules \\
\addlinespace[3pt]
Data flow & \texttt{defines}, \texttt{result\_of}, \texttt{DefUse} & Value origins, propagation, and instruction dependencies & All six vulnerability families \\
\addlinespace[3pt]
Memory access & \texttt{load}, \texttt{store}, \texttt{gep}, \texttt{ArrayAccess}, \texttt{PointerAccess} & Pointer dereferences, indexed accesses, and accessed objects & Buffer, array-bound, and invalid-pointer rules \\
\addlinespace[3pt]
Lifecycle & \texttt{alloc}, \texttt{free}, \texttt{ResourceAcquire}, \texttt{ResourceRelease} & Memory and resource acquisition-release relationships & Memory- and resource-leak rules \\
\addlinespace[3pt]
Calls and exits & \texttt{call}, \texttt{CallEdge}, \texttt{ExternalCallee}, \texttt{ErrorPath} & Inter-procedural transfers, external calls, and error exits & Lifecycle, pointer, and cross-family rules \\
\bottomrule
\end{tabular}
\end{table*}

\subsubsection{Deterministic IR-to-Fact Parsing}

The parser tokenizes instruction patterns, registers, operands, labels, constants, and types, then matches functions, basic blocks, instructions, and terminators and checks branch targets and instruction references.
Supported constructs are mapped to typed Base Facts, from which definition-use, reachability, control-dependence, memory-access, call, and error-path relations are constructed.
Before emitting facts, the parser checks and canonicalizes types, scopes, labels, register names, and equivalent instruction forms.
For available supporting functions selected in Stage~I, the parser connects actual arguments, formal parameters, return values, and call edges across function boundaries.
Unavailable callees are represented by \texttt{ExternalCallee} facts for the summaries supported by the Stage~III rule library.

\subsubsection{Fault Tolerance and Parallelization}

The parser marks unresolved fields \texttt{unknown}, isolates malformed instructions, retains parseable content, and logs excluded conflicting relations and structural coverage.
Missing positive premises prevent findings, potentially causing false negatives; safety conditions not derivable from available facts satisfy negated premises and may cause false positives when other conditions hold.
Derivation traces record positive facts and negated premises without a separate uncertain verdict; context gaps and structural coverage are recorded separately.
Non-derivability under recovered facts does not prove that the complete program lacks a safety condition.

Functions are parsed in parallel, and Stage~II facts are cached by the function identifier, canonicalized-IR SHA-256 hash, parser and schema revisions, and analysis-configuration hash.
Reuse requires an exact key match; Stage~III re-evaluates rules regardless of cache reuse.
The experiments in Section~\ref{SEC:EmpiricalStudy} run each configuration from an empty cache, so the reported runtimes do not include cache hits.

\subsection{Stage III: Explicit Vulnerability Rules and Datalog Execution}
\label{SEC:StageIII}

Stage~III evaluates explicitly defined vulnerability conditions over normalized evidence facts, with Souffl\'e Datalog serving as the execution engine.

\subsubsection{Rule Specification}

We organize the rule library around six vulnerability families, each associated with a representative CWE, as shown in Table~\ref{tab:vulnerability_rules}.
The four rule layers derive shared structural relations, identify positive operational evidence, recognize safety conditions and counterexamples, and apply decision rules and priorities to resolve overlapping interpretations.
Because the decision rules also require positive operational evidence, the absence of a safety fact alone cannot produce a vulnerability report.
The table presents schemas over Stage~II evidence and Stage~III auxiliary relations, omitting complete Souffl\'e type declarations.

Within the table, $A$ denotes the accessed array or memory object, $I$ the index, $B$ the associated basic block, $P$ a pointer, $R$ a resource, $E$ an exit block, $Op$ a division instruction, and $D$ its divisor.
\texttt{MayExceed}, \texttt{MayBeNull}, and \texttt{MayBeZero} express potential capacity, null-pointer, and zero-divisor violations supported by positive evidence.
\texttt{CompleteBoundsGuard} requires applicable \texttt{UpperGuard} and \texttt{LowerGuard} conditions for the same access, while \texttt{CapacityGuard}, \texttt{NullGuard}, and \texttt{NonZeroGuard} express the corresponding operation-specific protections.
\texttt{ReleasedOnPath(X,B,E)} represents release coverage along relevant acquisition-to-exit paths, while \texttt{PersistentReference(P)} denotes the memory-leak rule's retained-reference exception.
\texttt{Division(Op,D,B)} records a division instruction $Op$ with divisor $D$ in block $B$, and \texttt{Exit(E)} marks a basic block that terminates with a return; both are constructed mechanically from Base Facts in Stage~II.

\definecolor{rowgray}{gray}{0.95}
\begin{table*}[t]
\centering
\scriptsize
\caption{Representative decision rules.}
\label{tab:vulnerability_rules}
\setlength{\tabcolsep}{3pt}
\begin{tabular}{@{}>{\raggedright\arraybackslash}p{0.16\textwidth}>{\raggedright\arraybackslash}p{0.08\textwidth}>{\raggedright\arraybackslash}p{0.45\textwidth}>{\raggedright\arraybackslash}p{\dimexpr0.31\textwidth-6\tabcolsep\relax}@{}}
\toprule
\textbf{Vulnerability Type} &
\textbf{CWE} &
\textbf{Representative Decision Rule} &
\textbf{Key Evidence} \\
\midrule

\rowcolor{rowgray}
Buffer Overflow &
CWE-120 &
\ttfamily buffer\_overflow(A,I,B) :-
ArrayAccess(A,I,B),
MayExceed(I,A,B),
not CapacityGuard(I,A,B). &
\ttfamily ArrayAccess / MayExceed /
CapacityGuard \\

NULL Pointer Dereference &
CWE-476 &
\ttfamily null\_dereference(P,B) :-
PointerAccess(P,B),
MayBeNull(P,B),
not NullGuard(P,B). &
\ttfamily PointerAccess / MayBeNull /
NullGuard \\

\rowcolor{rowgray}
Memory Leak & 
CWE-401 & 
\ttfamily memory\_leak(P,E) :- 
ResourceAcquire(P,B), 
Reachable(B,E), Exit(E), 
not ReleasedOnPath(P,B,E), 
not PersistentReference(P). & 
\ttfamily ResourceAcquire / Reachable / Exit / ReleasedOnPath / PersistentReference \\

Missing Array Bounds Check &
CWE-129 &
\ttfamily missing\_bounds\_check(A,I,B) :-
ArrayAccess(A,I,B),
UpperGuard(I,A,B),
not CompleteBoundsGuard(I,A,B). &
\ttfamily ArrayAccess / UpperGuard /
CompleteBoundsGuard \\

\rowcolor{rowgray}
Division by Zero &
CWE-369 &
\ttfamily divide\_by\_zero(Op) :-
Division(Op,D,B),
MayBeZero(D,B),
not NonZeroGuard(D,B). &
\ttfamily Division / MayBeZero  /
NonZeroGuard \\

Resource Leak & CWE-772 & \ttfamily resource\_leak(R,E) :- ResourceAcquire(R,B), Reachable(B,E), Exit(E), not ReleasedOnPath(R,B,E). & \ttfamily ResourceAcquire / Reachable / Exit / ReleasedOnPath \\

\bottomrule
\end{tabular}
\end{table*}

\subsubsection{Path-, Dependency-, and Safety-Aware Reasoning}

The rule library recursively traverses definition-use relations and control-flow paths; operand provenance is computed as follows:

\begin{lstlisting}[
    language=prolog,
    basicstyle=\scriptsize\ttfamily,
    frame=single,
    breaklines=true
]
DependsOn(V, O) :-
    defines(I, V), operand(I, O).

DependsOn(V, O) :-
    defines(I, V), operand(I, U),
    DependsOn(U, O).
\end{lstlisting}

The resulting dependency relations are combined with control-dependence relations to determine whether a comparison guards the vulnerable operation on the relevant path.
An upper-bound comparison is insufficient when an index may be negative; a NULL check applies only if the checked value reaches the dereference without redefinition and the corresponding branch dominates that path.
For memory and resource lifecycles, cleanup operations must occur on every applicable acquisition-to-exit path.
Counterexample rules suppress otherwise unsupported reports when the available facts establish benign semantics, including persistent references, validated sanitizers, guaranteed initialization, or cleanup delegated to recognized APIs.
Priority rules resolve overlapping families, such as file-object versus generic pointer rules when file-lifecycle evidence is present, and buffer-capacity versus missing-bounds-check reports.

\subsubsection{Inter-Procedural and External-Call Rules}

Within the context selected in Stage~I, \texttt{CallEdge} facts and the actual-to-formal argument and return-value mappings emitted by the parser extend the recursive \texttt{DependsOn} relation across function boundaries, propagating value provenance and resource states.
This propagation is path-insensitive: it follows definition-use and call relations without tracking the branch conditions under which a call site is reached.
Consequently, these mappings alone do not establish a caller-side guard; suppression of a report requires an applicable safety condition to be derived from the recovered facts.
For recognized callees marked by ExternalCallee facts, Stage III applies explicit library-function summaries.
Unknown external calls are handled conservatively: existing taint or resource states may be propagated when supported by the rule, but an external call alone does not create a vulnerability verdict.

\subsubsection{Rule Execution and Reporting}

The engine verifies normalized fact types and arities and executes the rules using Souffl\'e~\cite{jordan2016souffle}.
Reports contain the CWE category, target function, source location, matched rule, supporting facts, and derivation trace.
Stable source identifiers pass from \texttt{source\_span} through IR \texttt{symlift.loc} annotations to \texttt{instr\_source} and \texttt{fact\_origin}.
Souffl\'e's explain-provenance interface provides each final tuple's proof tree; a manifest maps rule numbers to stable identifiers, and derived tuples inherit source identifiers from proof children.
The reported operation supplies the primary location, with other inherited locations retained as evidence.
Emission requires every positive proof branch to terminate at a source-backed program fact or explicitly identified rule-library fact, with a valid source anchor for the reported operation.
Trace completeness establishes neither full program recovery nor whole-program vulnerability reachability; fixed facts and rules yield repeatable verdicts.

\section{Empirical Study}
\label{SEC:EmpiricalStudy}

This section evaluates \SymLift through four research questions.

\subsection{Research Questions}

\begin{itemize}[leftmargin=*]
\item \textbf{RQ1: Compilation Gap Characterization.}
How does missing context affect compilation and program-relation recovery, and can \SymLift recover usable representations and complete analysis?
\item \textbf{RQ2: Vulnerability Detection Effectiveness.}
Does \SymLift outperform static analyzers, learned detectors, and direct LLMs on AI-generated code?
\item \textbf{RQ3: Component Contribution Analysis.}
How do representation choice, IR fidelity, generation and reasoning costs, and rule groups affect results?
\item \textbf{RQ4: Generalizability and Practical Applicability.}
How does \SymLift generalize to real CVE functions and context loss, and how do complexity and lifter choice affect effectiveness and cost?
\end{itemize}

\subsection{Experimental Setup}
\label{SEC:ExperimentalSetup}

\subsubsection{Datasets}

FormAI~\cite{formai}, Juliet~\cite{juliet}, and MegaVul~\cite{ni2024megavul} provide three complementary evaluation settings.
FormAI contains 1400 programs (1200 vulnerable, evenly divided among six families, and 200 non-vulnerable), comprising 4347 functions for IR analysis.
We form paired \emph{Juliet Full} and \emph{Juliet Fragment} sets from 600 programs and their extracted target functions.
MegaVul contains 1371 functions: 1171 CVE-related vulnerable functions and 200 non-vulnerable controls.

\textbf{Training Data.}
The structural-fidelity corpus contains 30,000 source-IR records: 25,167 function-IR records and 4833 structure-IR records.
The functions comprise 15,100 from Linux kernel 5.11.10, 5033 from OpenSSL 1.1.1k, and 5034 from SQLite 3.35.5.
All three use Ubuntu Clang 14.0.0-1ubuntu1.1/WLLVM 1.3.1 with \texttt{-O1 -fno-inline -g}.
For OpenSSL and SQLite, AST ranges and debug metadata align source functions with extracted IR definitions, excluding unrelated metadata and lifetime intrinsics.
Eligible pairs are selected by SHA-256 ranking of seed 42 concatenated with stable record identifiers.
Structure-IR pairs match named C structures to LLVM named types using Clang JSON AST and debug information, retaining dependencies and field order while excluding anonymous, forward-declared, opaque, or ambiguous matches; minimal IR modules are checked by \texttt{llvm-as}.

The vulnerability-oriented corpus contains 13,000 FormAI records: 7000 vulnerable functions selected using violation metadata and 6000 non-vulnerable functions sampled with seed 42.
Each has an unambiguous source body and reference IR accepted by \texttt{llvm-as} and the custom parser.
Deduplication hashes canonical source and non-derived Base Facts with SHA-256, excluding derived relations, vulnerability predicates, and labels.
Canonicalization removes comments and formatting, preserves source identifiers, literals, and statement order, and renumbers fact-side instruction, SSA, and block identifiers by first occurrence before sorting facts.
Label-consistent groups retain their lexicographically smallest record identifier; label-conflicting groups are excluded, removing 20 records and leaving 12,980.

\subsubsection{Training Configuration}
\label{SEC:TrainingConfiguration}

We fine-tune Qwen2.5-Coder-7B-Instruct using LoRA ($r=8$, $\alpha_{\mathrm{LoRA}}=16$, dropout 0.05) on all attention and MLP projections, with the base model frozen.
Training uses PyTorch/Transformers/PEFT, AdamW, cosine decay, warmup ratio 0.05, batch size 2, and eight gradient-accumulation steps.
Structural-feedback optimization uses learning rate $10^{-6}$ and jointly optimizes cross-entropy and policy-gradient terms with $\alpha=0.1$.
The structural-fidelity and vulnerability-oriented stages run for three and two epochs, respectively, using seed 42 for preparation and training; model selection follows Section~\ref{SEC:StageI}.

\subsubsection{Data Leakage Analysis}
\label{SEC:DataLeakage}

We assess vulnerability-oriented training--evaluation overlap using normalized source comparison and nearest-neighbor TF-IDF AST cosine similarity against 11,610 original training records.
AST-index exclusion does not affect training eligibility under source--IR alignment, reference-IR validation, and deduplication.
MegaVul coverage is 1165/1171 vulnerable functions (99.49\%) and 198/200 controls (99.00\%), totaling 1363/1371 (99.42\%); eight AST failures remain in detection.
Mean/95th-percentile similarities are 0.2345/0.3842 for vulnerable functions and 0.2210/0.3615 for controls.
Three vulnerable functions and no controls reach 0.80.
The same analysis of 4347 FormAI and 1200 paired Juliet functions yields mean/95th-percentile similarities of 0.4120/0.6385 and 0.1870/0.2940, respectively, with 12 FormAI functions (0.28\%) and no Juliet functions reaching 0.80.
No normalized exact duplicates are found in these comparisons.
Pairs reaching 0.80 are manually inspected and retained, not automatically deemed duplicates; the 12 FormAI pairs share input/output scaffolding but differ in vulnerability-relevant operations.
FormAI training and test sets share a generator, whereas Juliet is excluded from both training corpora.

\subsubsection{Compared Methods}
The static-analysis baselines are CodeQL~\cite{codeql}, Cppcheck~\cite{cppcheck}, and Clang Static Analyzer~\cite{clang-static-analyzer}.
The learned vulnerability-detection baselines are CodeBERT~\cite{feng2020codebert}, GraphCodeBERT~\cite{guo2020graphcodebert}, UniXcoder~\cite{guo2022unixcoder}, and CodeT5~\cite{wang2021codet5}.
The direct-LLM baselines are GPT-5.6-Terra~\cite{openai2026gpt56}, Claude-Opus-5~\cite{anthropic2026claudeopus5}, Gemini-3.5-Flash~\cite{google2026gemini35flash}, and DeepSeek-V4-Flash~\cite{xu2026deepseek}.

The same four models also serve as neural lifters, with their IR processed by the common \SymLift parser and Datalog backend.
Direct-Facts generates schema-constrained Base Facts without IR but retains relation construction, rules, and scoring.
Paired variants match context and generation budgets; \SymLift-Qwen-7B and its Direct-Facts variant additionally share the base checkpoint and aligned source records, supervised by IR or reference-derived Base Facts.

The IR-generation comparison includes \SymLift-Qwen-7B~\cite{qwen} and the four general-purpose \SymLift configurations, alongside the additional generators DeepSeek-R1-Distill-Qwen-14B~\cite{guo2025deepseek,deepseekai2025deepseekr1incentivizingreasoningcapability}, CodeLLaMA-13B~\cite{codellama}{\color{red}, }QwQ-LCoT-7B-Instruct~\cite{prithivmlmods2024qwqlcot7b}, and LLM-Compiler-13B~\cite{cummins2024meta}.

The MegaVul comparison additionally includes RepoAudit~\cite{guo2025repoaudit} as an academic auditing baseline, and Claude Code and Codex as industrial coding-agent baselines.

For a controlled comparison, each vulnerability-specific learned classifier is trained with 1000 positive samples from its target family and the same 1000 non-vulnerable samples.

Bare LLM names denote direct classification in detection tables; ``\SymLift-'' denotes IR-based configurations, including their IR-fidelity evaluations, and Direct-Facts marks their ablated variants.
Standalone IR generators remain unprefixed, with DeepSeek-R1-Distill-Qwen-14B and QwQ-LCoT-7B-Instruct abbreviated as \emph{DeepSeek-R1-14B} and \emph{QwQ-LCoT-7B}.

\subsubsection{Evaluation Protocol}
\label{SEC:EvaluationProtocol}

RQ1 evaluates representation availability and the \emph{analysis completion rate}, the proportion of inputs whose generated IR is parsed and normalized and whose Datalog execution completes.
Complete control-flow graph (CFG) and complete data-flow graph (DFG) rates are separate diagnostics, not completion prerequisites.

RQ2–RQ4 use \emph{Target-F1}, the standard F1 under benchmark-specific target matching, based on true positives (TP), false positives (FP), and false negatives (FN): $\mathrm{Target\mbox{-}F1}=2\mathrm{TP}/(2\mathrm{TP}+\mathrm{FP}+\mathrm{FN})$.
A true positive must match the target family and benchmark criterion; unmatched targets are false negatives, while wrong-family and non-vulnerable-sample alerts are false positives without cancelling correct predictions.
Technical failures remain in the fixed denominator and count as no-alert outputs; the false-negative rate (FNR) is defined as $\mathrm{FNR}=\mathrm{FN}/(\mathrm{TP}+\mathrm{FN})$.

FormAI sample-level target-matching accuracy (ACC) counts a vulnerable program as correct if any alert matches its target family and a non-vulnerable program only if no alert is reported; unlike Target-F1, it does not penalize additional wrong-family alerts.
MegaVul counts TP, FP, true negatives (TN), and FN over $1371\times6=8226$ sample--family decisions, pooling TP, FP, and FN for micro-averaged Target-F1.
Sample-level exact accuracy requires all six family decisions to match.

Representation ablation uses common reference-paired FormAI functions for fidelity and all FormAI programs for detection, unioning function alerts by program and deduplicating families.
CFG, DFG, and Base report control-flow, definition-use, and base-fact IoU; Overall covers all evaluated relations.
MegaVul reports Array Bounds (CWE-129), Buffer Overflow (CWE-120), Forgotten Memory/Resource (CWE-401 and CWE-772), Invalid Pointer (CWE-476), Division by Zero (CWE-369), and Invalid File-object Pointer (file-lifecycle priority specializations of CWE-476 and CWE-772).

Complexity uses five equally weighted features: comment-free token count, branch complexity, maximum control-structure nesting, call count, and pointer/array-operation count.
Branch complexity counts conditional statements, loop conditions, non-default switch cases, conditional expressions, and short-circuit logic; nesting covers conditionals, loops, and switches, while pointer/array operations include dereferences, address-taking, pointer-member accesses, and subscripts.
Each feature uses ascending ranks with averaged ties; with $r_{ij}$ denoting function $i$'s rank for feature $j$:
\begin{equation}
C_i = \frac{1}{5}\sum_{j=1}^{5}\frac{r_{ij}-1}{N-1},\qquad N=1371.
\label{eq:source-complexity}
\end{equation}
Sorting by $C_i$ and then stable sample identifier yields Low, Medium, and High tiers of 457 functions each, shared by all methods without using labels or outcomes.
The tiers are not family-balanced; comparisons are descriptive, not causal.

\subsection{RQ1: Compilation Gap Characterization}

RQ1 distinguishes surface parsing, independent compilation, and structural availability from successful downstream analysis.

\begin{table}[t]
\centering
\caption{Source-level representation availability across datasets.}
\label{tab:rq1-source-survival}
\scriptsize
\renewcommand{\arraystretch}{1.08}
\setlength{\tabcolsep}{1.2mm}
\begin{tabular}{llccccc}
\toprule
\textbf{Dataset} &
\textbf{N} &
\shortstack{\textbf{Tol.}\\\textbf{AST}} &
\shortstack{\textbf{Clang}\\\textbf{Succ.}} &
\shortstack{\textbf{CodeQL}\\\textbf{Func.}} &
\shortstack{\textbf{Nonempty}\\\textbf{CFG}} &
\shortstack{\textbf{Nonempty}\\\textbf{Def/Use}} \\
\midrule
FormAI
& 1400
& 100.00\%
& 100.00\%
& 100.00\%
& 100.00\%
& 99.93\% \\

Juliet Full
& 600
& 100.00\%
& 100.00\%
& 100.00\%
& 100.00\%
& 81.83\% \\

Juliet Fragment
& 600
& 100.00\%
& 12.00\%
& 100.00\%
& 100.00\%
& 65.83\% \\

MegaVul
& 1371
& 99.56\%
& 0.95\%
& 92.05\%
& 92.05\%
& 82.13\% \\
\bottomrule
\end{tabular}
\end{table}

Table~\ref{tab:rq1-source-survival} shows unchanged Juliet AST recovery despite compilation loss, and lower definition-use than CFG availability on MegaVul; operational analysis does not ensure all required relations are available.

\emph{Tol. AST} (tolerant AST recovery) requires Tree-sitter recovery of the target function and body without compilation or type resolution; \emph{Clang Succ.} (Clang compilation success) requires independent compilation.
CodeQL uses build-traced extraction (\texttt{manual}) for FormAI and Juliet Full and source-only extraction (\texttt{none}) for Juliet Fragment and MegaVul without restoring context.
\emph{CodeQL Func.} (CodeQL function availability), \emph{Nonempty CFG}, and \emph{Nonempty Def/Use} (nonempty definition-use relations) require a target body, at least one target-body CFG node, and at least one definition-use relation, respectively; extraction failures contribute zero under the full dataset denominator.
Juliet pairs match program identifiers and target-function names; target-body queries are deduplicated before counting.
Only 5.33\% and 47.67\% of pairs have equal CFG-node and definition-use counts, respectively.
The comparison measures sizes, not node correspondence: unequal counts establish divergence, but equal counts establish neither identical structure nor semantic equivalence; neither representation is ground truth.

Table~\ref{tab:rq1-symlift-survival} evaluates \SymLift-Qwen-7B: \emph{Model Succ.} (model success) requires nonempty IR without generation failure, and \emph{Target Func.} (target function) requires an identifiable target definition.
Complete CFGs require valid block terminators and resolved branch targets; complete DFGs require every use to resolve to a parameter, constant, or valid reaching definition.
These internal-consistency checks differ from RQ3's reference-based fidelity; all rates use full denominators, and partial representations may still complete analysis.

\begin{table}[t]
\centering
\caption{Structural availability and analysis completion of \SymLift-generated representations.}
\label{tab:rq1-symlift-survival}
\scriptsize
\renewcommand{\arraystretch}{1.08}
\setlength{\tabcolsep}{1mm}
\begin{tabular}{llccccc}
\toprule
\textbf{Dataset} &
\textbf{Unit} &
\shortstack{\textbf{Model}\\\textbf{Succ.}} &
\shortstack{\textbf{Target}\\\textbf{Func.}} &
\shortstack{\textbf{Complete}\\\textbf{CFG}} &
\shortstack{\textbf{Complete}\\\textbf{DFG}} &
\shortstack{\textbf{Analysis}\\\textbf{Compl.}} \\
\midrule
FormAI
& 4347 funcs.
& 99.52\%
& 98.60\%
& 75.87\%
& 77.64\%
& -- \\

Juliet Full
& 600 progs.
& 91.83\%
& 91.17\%
& 78.00\%
& 80.17\%
& 90.67\% \\

Juliet Fragment
& 600 progs.
& 91.00\%
& 90.33\%
& 75.67\%
& 80.33\%
& 89.50\% \\

MegaVul
& 1371 funcs.
& 100.00\%
& 80.16\%
& 48.14\%
& 49.31\%
& 76.37\% \\
\bottomrule
\end{tabular}
\end{table}

The analysis completion rate for FormAI is not reported because the function-level experiment evaluates structural availability rather than end-to-end vulnerability analysis.
Context removal reduces \SymLift's analysis completion rate on Juliet by only 1.17 percentage points, compared with an 88-percentage-point decrease in independent compilation.
On MegaVul, analysis completion exceeds both independent compilation and complete-CFG/DFG rates, showing that usable partial representations can support downstream analysis.

\textbf{Answer to RQ1.}
\SymLift enables vulnerability analysis without independent compilation; parsing success alone does not ensure usable semantic relations.

\subsection{RQ2: Vulnerability Detection Effectiveness}

Table~\ref{tab:rq2-formai} compares all methods on the fixed FormAI test set: \emph{Detected}, \emph{Missed}, and FNR use the 1200 vulnerable programs, whereas ACC and Target-F1 use all 1400 programs under the same target-matching protocol.

\begin{table}[t]
\centering
\caption{Vulnerability-detection effectiveness on FormAI. Det. and Miss. denote detected and missed vulnerable programs.}
\label{tab:rq2-formai}
\scriptsize
\renewcommand{\arraystretch}{1}
\setlength{\tabcolsep}{0.6mm}
\begin{tabular}{llccccc}
\toprule
\textbf{Category} &
\textbf{Method} &
\textbf{Det.} &
\textbf{Miss.} &
\textbf{ACC} &
\shortstack{\textbf{Target-F1}} &
\textbf{FNR} \\
\midrule
\multirow{3}{*}{Static analysis}
& CodeQL
& 198
& 1002
& 27.86\%
& 28.17\%
& 83.50\% \\

& Cppcheck
& 144
& 1056
& 24.36\%
& 21.38\%
& 88.00\% \\

& Clang Static Analyzer
& 120
& 1080
& 22.50\%
& 18.11\%
& 90.00\% \\

\midrule
\multirow{4}{*}{Learning-based}
& CodeBERT
& 571
& 629
& 48.79\%
& 61.43\%
& 52.42\% \\

& GraphCodeBERT
& 650
& 550
& 53.14\%
& 66.46\%
& 45.83\% \\

& UniXcoder
& 662
& 538
& 56.71\%
& 68.60\%
& 44.83\% \\

& CodeT5
& 546
& 654
& 45.21\%
& 58.74\%
& 54.50\% \\

\midrule
\multirow{4}{*}{Direct LLM}
& GPT-5.6-Terra
& 674 & 526 & 58.64\% & 69.95\% & 43.83\% \\

& Claude-Opus-5
& 701 & 499 & 57.64\% & 70.28\% & 41.58\% \\

& Gemini-3.5-Flash
& 685 & 515 & 58.71\% & 70.33\% & 42.92\% \\

& DeepSeek-V4-Flash
& 650 & 550 & 56.57\% & 68.13\% & 45.83\% \\

\midrule
\multirow{4}{*}{\SymLift}
& \textbf{\SymLift-Qwen-7B}
& \textbf{1123}
& \textbf{77}
& \textbf{90.50\%}
& \textbf{94.41\%}
& \textbf{6.42\%} \\

& \SymLift-GPT-5.6-Terra
& 921
& 279
& 67.43\%
& 80.16\%
& 23.25\% \\

& \SymLift-Claude-Opus-5
& 785
& 415
& 58.93\%
& 73.19\%
& 34.58\% \\

& \SymLift-Gemini-3.5-Flash
& 754
& 446
& 55.29\%
& 70.67\%
& 37.17\% \\

& \SymLift-DeepSeek-V4-Flash
& 758
& 442
& 55.43\%
& 70.84\%
& 36.83\% \\
\bottomrule
\end{tabular}
\end{table}

Static analyzers have low FormAI Target-F1 despite 100\% compilation success and high representation availability in RQ1; this comparison does not isolate rule coverage, target mapping, or relation quality.

\SymLift-Qwen-7B achieves the highest Target-F1, 94.41\%, exceeding the strongest direct-LLM and learned baselines by 24.08 and 25.81 percentage points, respectively.
All four general-purpose models improve when used for semantic lifting followed by symbolic reasoning rather than direct classification.
The largest gain is 10.21 percentage points for \SymLift-GPT-5.6-Terra over GPT-5.6-Terra.

\textbf{Answer to RQ2.}
\SymLift-Qwen-7B performs best on FormAI, highlighting the value of task-specific lifter training; general-purpose lifters improve here but not uniformly on MegaVul.

\subsection{RQ3: Component Contribution Analysis}

\textbf{Runtime Decomposition.}
Table~\ref{tab:rq3-latency} separates IR generation from symbolic reasoning for four \SymLift configurations on the same 200 FormAI samples.

\begin{table}[t]
\centering
\caption{Average runtime per FormAI sample, in seconds.}
\label{tab:rq3-latency}
\scriptsize
\renewcommand{\arraystretch}{1.08}
\setlength{\tabcolsep}{1.8mm}
\begin{tabular}{lcccc}
\toprule
\textbf{Configuration} &
\textbf{N} &
\textbf{Total} &
\textbf{IR Gen.} &
\textbf{Reasoning} \\
\midrule
\SymLift-Qwen-7B
& 200
& 6.793
& 6.415
& 0.378 \\

\SymLift-GPT-5.6-Terra
& 200
& \textbf{3.237}
& \textbf{2.834}
& 0.403 \\

\SymLift-Gemini-3.5-Flash
& 200
& 24.987
& 23.724
& 1.263 \\

\SymLift-DeepSeek-V4-Flash
& 200
& 10.628
& 9.979
& 0.649 \\
\bottomrule
\end{tabular}
\end{table}

Among the evaluated neural lifters, \SymLift-GPT-5.6-Terra achieves the shortest end-to-end (E2E) analysis time.
IR generation accounts for approximately 87.5\%--95.0\% of the total analysis time.

\begin{figure}[t]
    \centering
    \includegraphics[width=\columnwidth]{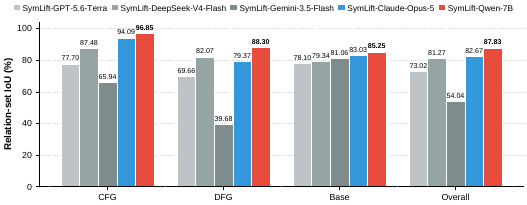}
    \caption{Intersection over union between generated and reference program relations on FormAI.}
    \label{fig:syminfer_decimal}
\end{figure}

\textbf{Representation Fidelity.}
Generated IR, directly generated facts, and reference IR are normalized into the same operation, branch, control-flow, and definition-use relations before computing relation-set intersection over union (IoU).
As shown in Figure~\ref{fig:syminfer_decimal}, \SymLift-Qwen-7B achieves 87.83\% overall IoU and 96.85\% control-flow IoU.

Table~\ref{tab:rq3-direct-facts} reports Direct-Facts variants on FormAI, with IR counterparts in Figure~\ref{fig:syminfer_decimal} and Table~\ref{tab:rq2-formai}.
All five \SymLift configurations achieve higher overall IoU and Target-F1 than their Direct-Facts variants; using Direct-Facts instead of IR for \SymLift-Qwen-7B reduces these metrics by 38.67 and 28.70 percentage points, respectively.
The benefit is not uniform across relations: the Direct-Facts variant of \SymLift-Gemini-3.5-Flash has higher base-feature and definition-use IoU than its IR counterpart.

\begin{table}[t]
\centering
\caption{Direct-Facts variants of the \SymLift configurations on FormAI. Parentheses show relative changes (\%) from IR--Facts.}
\label{tab:rq3-direct-facts}
\scriptsize
\renewcommand{\arraystretch}{1.08}
\setlength{\tabcolsep}{1.3pt}
\newcommand{\dfcell}[2]{\makebox[16pt][r]{#1}\hspace{0.5pt}{\fontsize{5.5}{6.5}\selectfont\makebox[19.2pt][l]{(#2)}}}
\newcommand{\dfname}[1]{\begin{tabular}[c]{@{}l@{}}\SymLift-\\#1\end{tabular}}
\begin{tabular}{@{}lrrrrr@{}}
\toprule
\multirow{2}{*}{\textbf{Configuration}} & \multicolumn{4}{c}{\textbf{Relation IoU (\%)}} & \multirow{2}{*}{\shortstack{\textbf{Target-F1}\\\textbf{(\%)}}} \\
\cmidrule(lr){2-5}
 & \multicolumn{1}{c}{\textbf{CFG}} & \multicolumn{1}{c}{\textbf{DFG}} & \multicolumn{1}{c}{\textbf{Base}} & \multicolumn{1}{c}{\textbf{Overall}} & \\
\midrule
\dfname{Qwen-7B} & \dfcell{55.93}{\textminus42.25} & \dfcell{44.17}{\textminus49.98} & \dfcell{78.92}{\textminus7.43} & \dfcell{49.16}{\textminus44.03} & \dfcell{65.71}{\textminus30.40} \\
\addlinespace[1pt]
\dfname{GPT-5.6-Terra} & \dfcell{40.58}{\textminus47.77} & \dfcell{33.81}{\textminus51.46} & \dfcell{70.90}{\textminus9.22} & \dfcell{35.93}{\textminus50.79} & \dfcell{39.02}{\textminus51.32} \\
\addlinespace[1pt]
\dfname{Claude-Opus-5} & \dfcell{34.25}{\textminus63.60} & \dfcell{32.84}{\textminus58.62} & \dfcell{58.22}{\textminus29.88} & \dfcell{33.00}{\textminus60.08} & \dfcell{37.78}{\textminus48.38} \\
\addlinespace[1pt]
\dfname{Gemini-3.5-Flash} & \dfcell{54.66}{\textminus17.11} & \dfcell{55.75}{+40.50} & \dfcell{81.55}{+0.60} & \dfcell{53.32}{\textminus1.33} & \dfcell{42.06}{\textminus40.48} \\
\addlinespace[1pt]
\dfname{DeepSeek-V4-Flash} & \dfcell{39.93}{\textminus54.36} & \dfcell{34.43}{\textminus58.05} & \dfcell{70.45}{\textminus11.20} & \dfcell{36.76}{\textminus54.77} & \dfcell{39.43}{\textminus44.34} \\
\bottomrule
\end{tabular}
\end{table}

We also report BLEU (Bilingual Evaluation Understudy), ROUGE-L (Recall-Oriented Understudy for Gisting Evaluation, Longest Common Subsequence), and METEOR (Metric for Evaluation of Translation with Explicit ORdering) as complementary text-similarity metrics in Table~\ref{tab:rq3-text-quality}.

\begin{table}[t]
\centering
\caption{Text similarity of generated intermediate representations.}
\label{tab:rq3-text-quality}
\scriptsize
\renewcommand{\arraystretch}{1.08}
\setlength{\tabcolsep}{2.0mm}
\begin{tabular}{lccc}
\toprule
\textbf{IR generator} &
\textbf{BLEU} &
\textbf{ROUGE-L} &
\textbf{METEOR} \\
\midrule
CodeLLaMA-13B
& 45.43\%
& 72.16\%
& 55.12\% \\

\textbf{{\SymLift-}Qwen-7B}
& \textbf{60.69\%}
& \textbf{81.75\%}
& \textbf{66.44\%} \\

LLM-Compiler-13B
& 51.14\%
& 76.33\%
& 59.37\% \\

QwQ-LCoT-7B
& 43.69\%
& 69.43\%
& 52.74\% \\

DeepSeek-R1-14B
& 48.54\%
& 74.27\%
& 57.12\% \\

\SymLift-DeepSeek-V4-Flash
& 49.86\%
& 75.67\%
& 58.31\% \\

\SymLift-GPT-5.6-Terra
& 30.12\%
& 65.34\%
& 43.73\% \\

\SymLift-Gemini-3.5-Flash
& 33.42\%
& 65.35\%
& 47.36\% \\
\bottomrule
\end{tabular}
\end{table}

\begin{figure}[t]
\centering
\includegraphics[width=\columnwidth]{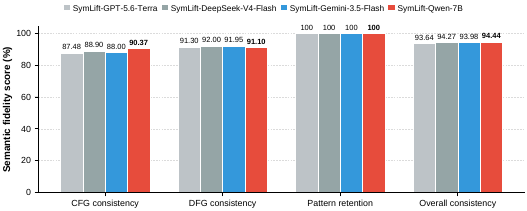}
\caption{Semantic-consistency inspection of generated IRs.}
\label{fig:manual_verification_unified_style}
\end{figure}

\SymLift-Qwen-7B leads all three textual metrics.
Manual inspection reports an overall consistency score of 94.44\% and 100\% pattern retention for \SymLift-Qwen-7B, as summarized in Figure~\ref{fig:manual_verification_unified_style}.
Relation IoU and manual semantic inspection provide the primary evidence of semantic fidelity because textual similarity alone cannot determine whether a definition-use or control-flow relation required by the backend is present.

\textbf{Rule-Group Ablation.}
Table~\ref{tab:rq3-component-ablation} reports the rule-group ablation results on MegaVul.To isolate the symbolic backend from generation variability, the MegaVul IR generated by \SymLift-Qwen-7B remains fixed while each configuration removes one symbolic rule group.
\emph{Drop} denotes the decrease in Target-F1, measured in percentage points, relative to the complete system.
\emph{Safety/Ctr./Priority}, \emph{Cross-fam.}, and \emph{Residual} denote combined safety/counterexample/priority rules, cross-family structural rules, and residual-completion rules, respectively.

\begin{table}[t]
\centering
\caption{Ablation of symbolic rule groups on MegaVul.}
\label{tab:rq3-component-ablation}
\scriptsize
\renewcommand{\arraystretch}{1.08}
\setlength{\tabcolsep}{1.1mm}
\begin{tabular}{lccccccc}
\toprule
\textbf{Configuration} &
\textbf{TP} &
\textbf{FP} &
\textbf{FN} &
\textbf{Precision} &
\textbf{Recall} &
\textbf{Target-F1} &
\textbf{Drop} \\
\midrule
\textbf{
\SymLift-Qwen-7B}
& \textbf{1006}
& \textbf{18}
& \textbf{165}
& \textbf{98.24\%}
& \textbf{85.91\%}
& \textbf{91.66\%}
& -- \\

w/o Safety/Ctr./Priority
& 1006
& 313
& 165
& 76.27\%
& 85.91\%
& 80.80\%
& \textbf{10.86} \\

w/o Cross-fam.
& 888
& 69
& 283
& 92.79\%
& 75.83\%
& 83.46\%
& 8.20 \\

w/o Lifecycle
& 872
& 29
& 299
& 96.78\%
& 74.47\%
& 84.17\%
& 7.49 \\

w/o Invalid ptr.
& 912
& 81
& 259
& 91.84\%
& 77.88\%
& 84.29\%
& 7.37 \\

w/o Capacity
& 915
& 47
& 256
& 95.11\%
& 78.14\%
& 85.79\%
& 5.87 \\

w/o Bounds
& 957
& 41
& 214
& 95.89\%
& 81.73\%
& 88.24\%
& 3.42 \\

w/o Residual
& 971
& 38
& 200
& 96.23\%
& 82.92\%
& 89.08\%
& 2.58 \\

w/o Div-zero
& 1004
& 45
& 167
& 95.71\%
& 85.74\%
& 90.45\%
& 1.21 \\

w/o File ptr.
& 995
& 18
& 176
& 98.22\%
& 84.97\%
& 91.12\%
& 0.55 \\
\bottomrule
\end{tabular}
\end{table}

The largest decrease, 10.86 percentage points, occurs when the combined safety, counterexample, and priority rules are removed.
TP remains 1006, but FP rises from 18 to 313, showing that these rules primarily reject unsupported alerts.
Removing cross-family structural rules causes the next-largest decrease, reducing TP and increasing FP.
Every evaluated rule-group removal reduces Target-F1 relative to the complete system.

\textbf{Answer to RQ3.}
IR generation dominates runtime; FormAI and fixed-IR ablations support IR-based lifting and explicit safety-aware vulnerability reasoning.

\subsection{RQ4: Generalizability and Practical Applicability}

RQ4 evaluates generalizability and full-workload cost.

\textbf{Real-World Vulnerability Detection.}
Table~\ref{tab:rq4-generalization} compares all methods on the complete MegaVul set under the target-matching protocol.
Each vulnerable function contributes one labeled target relation, whose recovery is recorded by \emph{Detected} and \emph{Missed}.

\begin{table}[t]
\centering
\caption{Vulnerability-detection effectiveness on MegaVul.}
\label{tab:rq4-generalization}
\scriptsize
\renewcommand{\arraystretch}{1.08}
\setlength{\tabcolsep}{1mm}
\begin{tabular}{llcccc}
\toprule
\textbf{Category} &
\textbf{Method} &
\textbf{Det.} &
\textbf{Miss.} &
\shortstack{\textbf{Target-}\\\textbf{F1}} &
\textbf{FNR} \\
\midrule

Static analysis
& CodeQL
& 77
& 1094
& 12.35\%
& 93.42\% \\

& Cppcheck
& 4
& 1167
& 0.68\%
& 99.66\% \\

\midrule
Learning-based
& CodeBERT
& 200
& 971
& 29.17\%
& 82.92\% \\

& GraphCodeBERT
& 122
& 1049
& 18.87\%
& 89.58\% \\

& UniXcoder
& 160
& 1011
& 24.04\%
& 86.34\% \\

& CodeT5
& 199
& 972
& 29.03\%
& 83.01\% \\

\midrule
Academic hybrid
& RepoAudit
& 42
& 1129
& 6.92\%
& 96.41\% \\

\midrule
Industrial agent
& Claude Code
& 354
& 817
& 46.43\%
& 69.77\% \\

& Codex
& 348
& 823
& 45.82\%
& 70.28\% \\

\midrule
Direct LLM
& GPT-5.6-Terra
& 256
& 915
& 26.27\%
& 78.14\% \\

& Claude-Opus-5
& 339
& 832
& 28.45\%
& 71.05\% \\

& Gemini-3.5-Flash
& 333
& 838
& 30.01\%
& 71.56\% \\

& DeepSeek-V4-Flash
& 318
& 853
& 30.00\%
& 72.84\% \\

\midrule
\SymLift
& \textbf{\SymLift-Qwen-7B}
& \textbf{1006}
& \textbf{165}
& \textbf{91.66\%}
& \textbf{14.09\%} \\

& \SymLift-GPT-5.6-Terra
& 183
& 988
& 26.60\%
& 84.37\% \\

& \SymLift-Claude-Opus-5
& 209
& 962
& 29.71\%
& 82.15\% \\

& \SymLift-Gemini-3.5-Flash
& 169
& 1002
& 24.85\%
& 85.57\% \\

& \SymLift-DeepSeek-V4-Flash
& 172
& 999
& 25.26\%
& 85.31\% \\

\bottomrule
\end{tabular}
\end{table}

\SymLift-Qwen-7B's 91.66\% Target-F1 exceeds the strongest direct-LLM baseline, Gemini-3.5-Flash, by 61.65 percentage points and the strongest non-\SymLift baseline, Claude Code, by 45.23 percentage points.
The same configuration obtains 98.24\% micro-averaged precision and 87.02\% sample-level exact accuracy.
A 10,000-round repository-clustered bootstrap gives a 95\% confidence interval of [88.65\%, 93.69\%] for Target-F1.

Compared with direct source classification, \SymLift-GPT-5.6-Terra and \SymLift-Claude-Opus-5 improve slightly in Target-F1, whereas \SymLift-Gemini-3.5-Flash and \SymLift-DeepSeek-V4-Flash decline.
Unlike the consistent improvements observed on FormAI, the MegaVul results do not show a uniform benefit across the evaluated model families.
The four general-purpose lifters have precision of 88.56\%--90.05\% but recall of only 14.43\%--17.85\%.
All four recover fewer targets than their direct-classification counterparts and remain below the fine-tuned \SymLift-Qwen-7B configuration in Target-F1.

\textbf{Controlled Context Loss.}
Table~\ref{tab:rq4-juliet-direct} compares \SymLift-Qwen-7B with four direct-LLM baselines on Juliet Full and Fragment.
A correct detection requires matching both the vulnerability family and statement; incorrect localization counts as FP and leaves the target unmatched.

\begin{table}[t]
\centering
\caption{Vulnerability-detection effectiveness on Juliet Full and Fragment.}
\label{tab:rq4-juliet-direct}
\scriptsize
\renewcommand{\arraystretch}{1.08}
\setlength{\tabcolsep}{1.8mm}
\begin{tabular}{llccc}
\toprule
\textbf{Method} &
\textbf{Input} &
\textbf{Precision} &
\textbf{Recall} &
\textbf{Target-F1} \\
\midrule
GPT-5.6-Terra
& Full
& 64.84\%
& 67.00\%
& 65.90\% \\

& Fragment
& 62.97\%
& 66.33\%
& 64.61\% \\

Claude-Opus-5
& Full
& 64.69\%
& 76.33\%
& 70.03\% \\

& Fragment
& 62.26\%
& 77.00\%
& 68.85\% \\

Gemini-3.5-Flash
& Full
& 77.38\%
& 78.67\%
& 78.02\% \\

& Fragment
& 81.94\%
& 84.67\%
& 83.28\% \\

DeepSeek-V4-Flash
& Full
& 91.26\%
& \textbf{94.00\%}
& 92.61\% \\

& Fragment
& 85.40\%
& 89.67\%
& 87.48\% \\

\midrule
\textbf{\SymLift-Qwen-7B}
& Full
& \textbf{94.48\%}
& 91.33\%
& \textbf{92.88\%} \\

& Fragment
& \textbf{94.54\%}
& \textbf{92.33\%}
& \textbf{93.42\%} \\
\bottomrule
\end{tabular}
\end{table}

\SymLift-Qwen-7B achieves the highest Target-F1 under both context conditions, outperforming the strongest direct LLM, DeepSeek-V4-Flash, by 5.94 percentage points on Juliet Fragment.
The 0.54-percentage-point change indicates stable aggregate detection under the evaluated context-removal procedure, despite the compilation decline reported in RQ1.

\begin{table}[t]
\centering
\caption{Per-family performance of \SymLift-Qwen-7B on MegaVul.}
\label{tab:rq4-category}
\scriptsize
\renewcommand{\arraystretch}{1.08}
\setlength{\tabcolsep}{2mm}
\begin{tabular}{lcccccc}
\toprule
\textbf{Family} &
\textbf{TP} &
\textbf{FP} &
\textbf{FN} &
\textbf{Precision} &
\textbf{Recall} &
\textbf{Target-F1} \\
\midrule
Array Bounds
& 180
& 14
& 20
& 92.78\%
& 90.00\%
& 91.37\% \\

Buffer Overflow
& 200
& 1
& 0
& 99.50\%
& 100.00\%
& \textbf{99.75\%} \\

\shortstack[l]{Forgotten Memory/\\Resource}
& 185
& 0
& 15
& 100.00\%
& 92.50\%
& 96.10\% \\

Invalid Pointer
& 197
& 0
& 3
& 100.00\%
& 98.50\%
& 99.24\% \\

Division by Zero
& 91
& 3
& 109
& 96.81\%
& 45.50\%
& 61.90\% \\

\shortstack[l]{Invalid File-object\\Pointer}
& 153
& 0
& 18
& 100.00\%
& 89.47\%
& 94.44\% \\
\bottomrule
\end{tabular}
\end{table}

Table~\ref{tab:rq4-category} evaluates each MegaVul family on the same 1371-sample set.
Five evaluation families exceed 90\% Target-F1.
Division by Zero remains the main limitation at 61.90\%.
Its missed cases commonly require divisor propagation through parameters, aliases, or relations omitted during IR generation, consistent with the failure analysis in Section~\ref{SEC:Discussion}.

\textbf{Impact of Code Complexity.}
Table~\ref{tab:rq4-complexity} compares the Target-F1 of five \SymLift configurations across the same three source-complexity tiers of 457 samples each.

\begin{table}[t]
\centering
\caption{Target-F1 across source-code complexity tiers on MegaVul.}
\label{tab:rq4-complexity}
\scriptsize
\renewcommand{\arraystretch}{1.08}
\setlength{\tabcolsep}{2.5mm}
\begin{tabular}{lcccc}
\toprule
\textbf{\SymLift Configuration} &
\textbf{Low} &
\textbf{Medium} &
\textbf{High} &
\textbf{Overall} \\
\midrule

\textbf{\SymLift-Qwen-7B}
& \textbf{90.08\%}
& \textbf{90.98\%}
& \textbf{93.44\%}
& \textbf{91.66\%} \\

\SymLift-GPT-5.6-Terra
& 18.83\%
& 19.18\%
& 38.17\%
& 26.60\% \\

\SymLift-Claude-Opus-5
& 21.89\%
& 18.26\%
& 44.09\%
& 29.71\% \\

\SymLift-Gemini-3.5-Flash
& 18.37\%
& 16.36\%
& 36.30\%
& 24.85\% \\

\SymLift-DeepSeek-V4-Flash
& 16.54\%
& 16.32\%
& 38.46\%
& 25.26\% \\

\bottomrule
\end{tabular}
\end{table}

\SymLift-Qwen-7B achieves the highest Target-F1 in every tier, maintaining scores of 90.08\%--93.44\%.
The four general-purpose lifters score higher in the High tier than in the Low or Medium tiers, but remain below \SymLift-Qwen-7B throughout.
These comparisons do not establish a causal effect of source-code complexity.

\textbf{Runtime and Token Cost.}
Table~\ref{tab:rq4-cost} compares the cost of processing all 1371 MegaVul samples by direct source classification and the \SymLift pipeline.
Request-level telemetry provides runtime and input/output token counts for the four direct-LLM baselines and five neural lifters; summed token counts measure inference volume rather than monetary price.

\begin{table}[t]
\centering
\caption{Runtime and token cost on MegaVul. Runtime entries are total hours/average seconds.}
\label{tab:rq4-cost}
\scriptsize
\renewcommand{\arraystretch}{1.08}
\setlength{\tabcolsep}{0.45mm}
\begin{tabular}{lcccc}
\toprule
\textbf{Configuration} &
\shortstack{\textbf{In/Out}\\\textbf{Tokens}} &
\shortstack{\textbf{Model}\\\textbf{Time (h/s)}} &
\shortstack{\textbf{Backend}\\\textbf{Time (h/s)}} &
\shortstack{\textbf{E2E}\\\textbf{Time (h/s)}} \\
\midrule
Direct GPT-5.6-Terra
& 1.51M/0.57M
& 8.02/21.05
& --
& 8.02/21.05 \\

Direct Claude-Opus-5
& 2.41M/3.30M
& 17.30/45.43
& --
& 17.30/45.43 \\

Direct Gemini-3.5-Flash
& 2.07M/6.13M
& \textbf{6.79/17.84}
& --
& \textbf{6.79/17.84} \\

Direct DeepSeek-V4-Flash
& 1.71M/19.59M
& 51.03/133.98
& --
& 51.03/133.98 \\

\midrule
\textbf{\SymLift-Qwen-7B}
& 1.13M/3.33M
& 21.48/56.39
& 0.33/1.13
& \textbf{21.81/57.26} \\

\SymLift-GPT-5.6-Terra
& 1.12M/5.17M
& 31.52/82.76
& 1.18/3.10
& 32.70/85.85 \\

\SymLift-Claude-Opus-5
& 8.04M/6.11M
& 39.95/104.90
& 1.19/3.15
& 41.14/108.01 \\

\SymLift-Gemini-3.5-Flash
& 1.62M/22.68M
& 23.60/61.97
& 1.24/3.29
& 24.84/65.23 \\

\SymLift-DeepSeek-V4-Flash
& 1.32M/27.12M
& 63.65/167.13
& 1.23/3.32
& 64.88/170.37 \\
\bottomrule
\end{tabular}
\end{table}

Among direct baselines, Gemini-3.5-Flash is fastest, while GPT-5.6-Terra uses the fewest tokens (2.08 million).
Among \SymLift configurations, \SymLift-Qwen-7B is fastest and uses the fewest tokens (4.46 million).
Most cross-model time variation comes from neural generation, with the backend averaging 1.13--3.32 seconds per successful execution.
Model and end-to-end averages use all 1371 inputs, whereas backend averages use successful backend executions.

\textbf{Answer to RQ4.}
\SymLift-Qwen-7B remains effective across MegaVul, Juliet contexts, and complexity tiers; general-purpose lifters are recall-limited, while IR generation dominates runtime.

\section{Discussion}
\label{SEC:Discussion}

This section interprets the empirical findings and examines the practical scope of \SymLift.

\subsection{Practical Implications}

In IDE-based code generation, function auditing, and code review, \SymLift enables analysis without full-build reconstruction by recovering sufficiently faithful local semantics.

\textbf{Representation Fidelity.}
The MegaVul lifter and FormAI Direct-Facts comparisons motivate evaluating relation fidelity and downstream detection, not generation coverage alone.

\textbf{Independent Optimization.}
IR-generation cost motivates quantization, distillation, caching, incremental generation, or efficient decoding while retaining the rule library, provided that IR satisfies the parser's contract.

\subsection{Failure Cases}

The three cases come from the fixed 1371-sample MegaVul set; their zero-padded evaluation-list identifiers map to CVEs, repositories, fixing commits, and target functions in \texttt{failure\_cases/manifest.csv}.
The replication package provides source fragments, generated IR, parsing logs, facts, and derivation or failure reports under \texttt{failure\_cases/<identifier>/}.

\textbf{Output Truncation.}
In one division-by-zero case, a large protocol-parsing routine's generated IR reaches the 8192-token limit with \texttt{finish\_reason=length}, omitting \texttt{ret} and all \texttt{sdiv}, \texttt{udiv}, and \texttt{fdiv} instructions.
The parser retains the parseable prefix, but the missing division prevents a \texttt{Division} fact and causes a false negative originating in semantic lifting rather than symbolic reasoning.

\textbf{Missing Inter-Procedural Resource Semantics.}
In another case (labeled Forgotten Memory/Resource), generated IR preserves resource-management calls whose callees lie outside the selected context.
They are represented as \texttt{ExternalCallee} facts without matching summaries, so their resource effects are unavailable and \texttt{ReleasedOnPath} cannot be established for acquired descriptors.
This failure motivates extending the relevant external-call summaries without reconstructing the complete project.

\textbf{Collapsed Pointer Definitions.}
A third case (labeled Invalid Pointer) checks a lookup result against NULL, then reassigns the pointer from an unchecked second call before dereferencing it on a cleanup path.
The generated IR omits the second call and collapses both definitions into one SSA value.
The parser connects the checked value to the dereference through \texttt{DefUse}; however, although the rules establish \texttt{NullGuard}, the absence of \texttt{MayBeNull} evidence
for the unchecked value prevents derivation of \texttt{null\_dereference}, resulting in a false negative attributable to the loss of the redefinition and its
unchecked producer during semantic lifting.

\subsection{Limitations}

\textbf{IR Generation Uncertainty.}
Generated IR can still omit instructions, recover incorrect types, corrupt definition-use relations, or be truncated; deterministic Datalog execution does not remove this dependence on semantic fidelity.
Stronger structural validation and uncertainty estimation therefore remain important directions.

\textbf{Function-Level Context.}
The bounded context selected in Section~\ref{SEC:StageI} may omit relations involving
unavailable implementations, unresolved indirect calls, or excluded functions; aliases, callbacks, and global resource lifecycles remain challenging when available context or summaries are insufficient.

\textbf{Vulnerability Scope.}
Our current rules cover six representative vulnerability families, not the complete vulnerability space.
Extending \SymLift to additional CWE families, concurrency bugs, injection vulnerabilities, and higher-level logic errors will require new facts and reasoning rules.

\textbf{Evaluation and Cost.}
Existing hybrid approaches differ in languages, granularity, build requirements, and output semantics, limiting uniform comparison.
Although symbolic reasoning is inexpensive, IR-generation cost must be reduced before claiming low-cost, large-scale real-time deployment.

\section{Related Work}
\label{SEC:Related}

Vulnerability detection has been explored through static program analysis, learning-based models, LLM-based evaluation, and hybrid reasoning systems~\cite{cui2026novel,yi2025adaptive}.

\textbf{Static Analysis Tools}.
Tools such as CodeQL~\cite{codeql}, Clang Static Analyzer~\cite{clang-static-analyzer}, and Cppcheck~\cite{cppcheck} reason over structured program representations.
Lightweight syntactic matchers, e.g., Semgrep~\cite{Semgrep2021}, reduce compilation dependencies.
Empirical evaluations report both false positives and missed defects; combining tools can improve coverage without eliminating these limitations~\cite{lipp2022empirical}.
For build-dependent analysis configurations, missing declarations, types, or dependencies can prevent representation construction.

\textbf{Learning-based Detection}.
Learning-based detection includes VulDeePecker~\cite{li2018vuldeepecker} and downstream use of CodeBERT~\cite{feng2020codebert}, GraphCodeBERT~\cite{guo2020graphcodebert}, and UniXcoder~\cite{guo2022unixcoder}.
VulPrompt~\cite{irtiza2024vulprompt} explores few-shot prompt-based graph learning.
However, a learned classification score alone does not establish a derivation from explicit program relations.
Actionable evidence depends on explanation and localization~\cite{sovrano2025large}, while evaluation must account for dataset construction~\cite{daneshvar2026vulscriber,yi2026exploring}, code characteristics~\cite{weissberg2026llm}, and representativeness~\cite{fan2020ac}.

\textbf{LLM-based Approaches}.
Recent work studies LLM-generated code~\cite{mo2025assessing,zhuo2025bigcodebench,zhuo2026identifying,spracklen2025we} and applies LLMs directly to security tasks~\cite{nazzal2024promsec,li2026exploratory,el2026repairing,wang2026evaluation,wu2025mystique}.
A systematic literature review covers their broader software-engineering applications~\cite{Hou2023_LLM4SE}.
Zhang et al.~\cite{zhang2024prompt} studied prompt designs for ChatGPT-based vulnerability detection.
LLMs have also been adapted for bug finding~\cite{wang2025securemind} and decompilation~\cite{tan2024llm4decompile}, but generated explanations require independent validation before serving as program evidence.

\textbf{Hybrid Reasoning Systems}.
Hybrid approaches combine symbolic program analysis with learning- or LLM-based components~\cite{wang2026nesa,wang2024llmdfa,lekssays2025llmxcpg,ibrahimzada2025alphatrans,granberry2026seeking,mohammed2025llm}.
Early systems such as VulDeePecker~\cite{li2018vuldeepecker} and SySeVR~\cite{li2021sysevr} encode vulnerability-relevant patterns for neural classifiers, while Devign applies GNNs to program graphs~\cite{zhou2019devign}.
More recent approaches use LLMs to infer analysis specifications~\cite{shemetova2025lamed} or synthesize static-analysis queries, as in IRIS~\cite{li2024iris} and MoCQ~\cite{li2025automated}.
Other efforts pair generated artifacts with fuzzing or assertion checks~\cite{wang2024sanitizing,yao2024fuzzllm, pulavarthi2024assertionbench, tian2025assertcoder}.
IRIS's Java analysis follows project compilation and CodeQL database construction~\cite{li2024iris}.
MoCQ executes queries over Joern or CodeQL representations~\cite{li2025automated}.
Both depend on frontend-recovered relations, with compilation requirements varying by frontend and language.
RepoAudit~\cite{guo2025repoaudit} performs compilation-free repository auditing through on-demand exploration and LLM-inferred data-flow facts.
Its validation combines parsing-based control-flow checks with LLM-based interprocedural path checks.
The comparison with RepoAudit, LLMDFA, and NESA~\cite{guo2025repoaudit,wang2024llmdfa,wang2026nesa} therefore concerns evidence construction and vulnerability judgment, not compilation freedom or Datalog alone.
\SymLift assigns LLVM-like IR reconstruction from context-incomplete fragments to the neural lifter and executes explicit vulnerability rules using Souffl\'e~\cite{jordan2016souffle}.

\section{Threats to Validity}
\label{SEC:ThreatsToValidity}

The current study is subject to several validity threats.
First, IR quality is measured using fact-level IoU, text-similarity metrics, and manual inspection, which may not capture full behavioral equivalence.
Second, parser and rule defects or assumptions may influence the results.
We reduce this risk through syntax validation, fact normalization, held-out evaluation, and manual verification.
Third, evaluation on C code, six vulnerability families, FormAI, paired Juliet Full and Fragment inputs, MegaVul, and selected baselines may not generalize to other languages, projects, or families.
Finally, tool--input combinations outside the applicable language or analysis mode are unsupported and receive no numerical detection score; evaluated technical failures remain no-alert outputs in the fixed denominator, not unsupported configurations.
A no-alert output caused by failure does not establish that the input is safe.

Within each benchmark condition, methods receive the same permitted source context, with Full and Fragment evaluated separately; settings, prompts, and budgets are fixed per configuration rather than adjusted to test outcomes.
Nevertheless, differences in supported analyses and resource requirements limit the interpretation of cross-method comparisons.

Source-level duplicate and AST-similarity checks (Section~\ref{SEC:DataLeakage}) do not exclude all semantic similarity or foundation-model pretraining overlap.
Manual semantic inspection remains susceptible to reviewer judgment and expectations; it is not treated as a proof of behavioral equivalence.

\section{Conclusion}
\label{SEC:Conclusion}

This paper presented \SymLift, a neuro-symbolic framework for detecting vulnerabilities in context-incomplete code fragments.
By separating neural semantic lifting from rule-based inference, \SymLift supports compile-free operation at analysis time and deterministic vulnerability reasoning for fixed evidence facts and rules.
SymLift-Qwen-7B achieves 94.41\% Target-F1 on FormAI and 91.66\% micro-averaged Target-F1 on MegaVul, with only a 0.54-percentage-point Target-F1 change under controlled Juliet context removal.
Neural-generation errors, incomplete context, and generation cost remain limitations; future work will reduce latency and extend CWE coverage.

\balance
\bibliographystyle{IEEEtran}
\bibliography{IEEEabrv,references}
\end{document}